%% file: main.tex
\documentclass[11pt]{article}
\usepackage[left=1in,right=1in,top=1in,bottom=1in]{geometry}
\usepackage[T1]{fontenc}
\usepackage{lmodern,microtype}
\IfFileExists{times.sty}{\usepackage{times}}{}
\usepackage{amsmath,amssymb,amsthm,mathtools,aliascnt}
\usepackage[cal=stixtwofancy]{mathalpha}
\usepackage{booktabs,longtable,array,needspace,graphicx,float}
\usepackage{hyperref,cleveref}
\usepackage[authordate,backend=biber,maxcitenames=2,uniquename=false]{biblatex-chicago}
\hypersetup{colorlinks=true,linkcolor=blue,citecolor=blue,urlcolor=blue}
\newtheorem{proposition}{Proposition}
\newtheorem{lemma}{Lemma}[section]
\theoremstyle{definition}
\newtheorem{assumption}{Assumption}[section]
\newaliascnt{regularity}{assumption}
\newtheorem{regularity}[regularity]{Regularity condition}
\aliascntresetthe{regularity}

\theoremstyle{remark}
\crefname{proposition}{proposition}{propositions}\Crefname{proposition}{Proposition}{Propositions}
\crefname{lemma}{lemma}{lemmas}\Crefname{lemma}{Lemma}{Lemmas}
\crefname{assumption}{assumption}{assumptions}\Crefname{assumption}{Assumption}{Assumptions}
\crefname{regularity}{regularity condition}{regularity conditions}\Crefname{regularity}{Regularity condition}{Regularity conditions}
\crefname{definition}{definition}{definitions}\Crefname{definition}{Definition}{Definitions}
\newcommand{\R}{\mathbb R}\newcommand{\E}{\mathbb E}
\newcommand{\Var}{\operatorname{Var}}\newcommand{\Cov}{\operatorname{Cov}}
\newcommand{\tr}{\operatorname{tr}}

\newcommand{\pto}{\xrightarrow{p}}\newcommand{\dto}{\Rightarrow}

\DeclareMathOperator*{\argmin}{arg\,min}
\newcommand{\proofref}[1]{\par\noindent\emph{Proof:} Appendix~\ref{#1}.}
\newcommand{\statementend}{\unskip\nobreak\hfill\ensuremath{\triangle}}
\author{Christopher W. Engh\\ \textit{Yale University}}
\date{}
\title{Exploiting Locally Exogeneous Variation in Spatial Data:\\
Linear Regression with Many Small Units Using InfillGLS\footnote{Preliminary draft. An associated \texttt{R} package and proof formalizations (including those in Lean) were completed by \texttt{GPT 5.6}. These will be located at \texttt{github.com/chrisw2k/infillGLS/}. The author thanks Yuichi Kitamura, Tim Christensen, Leah Boustan, and his fellow Yale PhD students for helpful discussions. All errors in the manuscript, code, and formalization are the sole responsibility of the author.}}
\begin{document}
\maketitle
\vspace{-0.3in}
\begin{abstract}
    This article shows how researchers can exploit \textit{locally exogeneous variation} even in the presence of endogeneity, autocorrelation, or spurious correlation. It describes the infill asymptotic properties of generalized least squares estimation with county-level data, Census tract-level data, and similar regional averages, including sufficient conditions for consistency, asymptotic normality, and valid inference. The \texttt{R} package \texttt{infillGLS} implements our methods.
\end{abstract}
\begin{figure}[H]
\centering
\includegraphics[width=\textwidth]{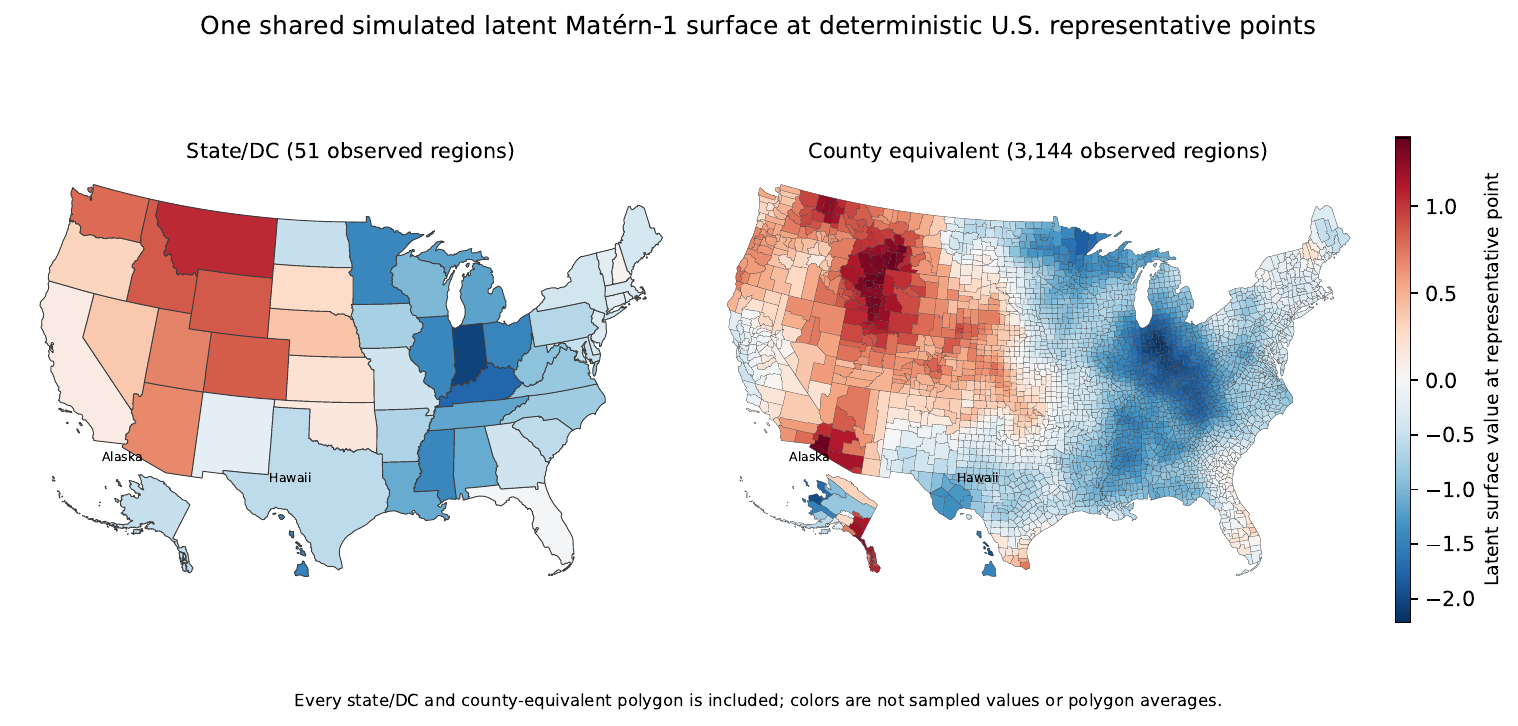}
\caption{One shared simulated latent Mat\'ern-$1$ surface evaluated at
deterministic representative points for every observed U.S.~state/DC and
county-equivalent polygon.  The panels contain all $51$ state/DC regions and
all $3{,}144$ county-equivalents.}
\label{fig:us-geometry-surface}
\end{figure}

\section{Introduction}

Spatial econometric datasets grow in size in the manner illustrated in \cref{fig:us-geometry-surface}: by partitioning the same domain (e.g. the United States) into increasingly fine regional units (states, counties, tracts, ...). This eminently intuitive regime is known as \textit{infill} or {fixed domain asymptotics}. 

This article describes when and how a researcher can identify, estimate, and conduct inference on the slope parameter of a linear model by exploiting \textit{locally exogenous variation}---a term we use to describe conditions under which finer regional data provide increasing treatment information after controls while the conditional mean of the error given the regressors makes a negligible contribution to the slope estimate. These conditions permit treatment to be endogenous in levels and errors to remain spatially correlated, including at fine resolutions. Under the consistency assumptions, the conditional slope bias tends to zero; under the inference assumptions, it is negligible relative to the slope's standard deviation. Heuristically, this allows for identification through a spatial analog of time series first differences (although in practice, we will not be literally first differencing).

The aforementioned intuition behind identification via local exogeneity has been recognized throughout the social sciences, usually taking on a stronger form which states that \textit{conditional} on a super-region (e.g. a state), the treatment assigned to its sub-regions (e.g. counties) is as-good-as-random. For example, \textcite{algan2016social} study the effects of ethnic diversity within city blocks in France, which they identify by arguing that sorting on diversity occurs on the neighbourhood level, but not on the block level. \textcite{bayer2025distinguishing} study neighbourhood racial change (``white flight'') using data on an even finer scale: they identify the effect of a next-door neighbour of a different race by controlling for the characteristics of neighbours two or three doors down. \textcite{lyall2009does} studies the effects of indiscriminate shelling in Chechnya, which he identifies by arguing that although the allocation of artillery pieces to an area was endogeneous but being hit by a shell conditional on being within range of artillery was random. In economic history and political economy, a cottage industry has investigated the effect of exposure to various radio and television programs by appealing to highly plausible sources of exogenous local variation, such as topographical features \parencite{yanagizawa2014propaganda,dellavigna2014cross,durante2019political}. 

However, the empirical strategies employed in these settings have been varied and \textit{ad hoc}, lacking a unified framework for understanding whence the asymptotic properties of the resulting estimates actually come from. The identifying assumptions prescribed are often stronger than need be, and the estimation procedures often rely on clunky regional fixed effects which abstract away the continuous nature of spatial confounding.

This article formalizes the intuition behind local exogeneity and the asymptotic properties of generalized least-squares (GLS) regression. Treatment may be endogenous in levels and autocorrelated. Consistency requires mild regularity conditions, namely: useful treatment variation after controls, a uniform bound on true variances relative to working variances, and a restriction on the conditional mean of the error given the regressors. The methods permit nonzero confounding and a misspecified working covariance. Asymptotic normality and valid inference require stronger regularity conditions controlling the degree of misspecification and on quadratic forms of the disturbance surface. 

\noindent \textit{How does an infill model diverge from traditional spatial econometrics?} The vast, vast majority spatial econometric models use expanding domain asymptotics. The difference lies in where new information is assumed to come from as $n\to\infty$. The common problem is non-independence: additional observations may be correlated with existing observations. Expanding domain asymptotics typically rely on mixing conditions of some sort, whereby dependence between units vanishes with distance; as more units are observed, each additional unit---by virtue of being far from some existing units---contributes additional information both in levels and in differences. By contrast, infill asymptotics relies on the existence of local differences at finer spatial scales; as the researcher ``zooms in'' on a map, they should be able to observe local variation in treatment and compare that with local variation in outcomes. Thus, as the number of units increase, additional information comes from drawing new local contrasts that were previously hidden at coarser levels of aggregation.

\noindent \textit{Related literature.} The covariance-space approach to fixed-domain regression originates in \textcite{grenander1950,parzen1961,parzen1962}. \textcite{guan2023} describe spatial confounding through frequency-specific dependence. \textcite{gilbert2025} establish GLS consistency under a spatial-confounding model with nonspatial exposure variation. \textcite{dattastein2026} characterize consistent estimability between random fields and construct local difference-based estimators. These results motivate a distinction between available treatment information, restrictions on confounding, and validity of a particular regression weighting.

Feasible inference under infill and local exogeneity also has high-frequency precedents. \textcite{litodorovtauchen2017} develop optimally weighted inference for a constant continuous-time regression coefficient, and \textcite{aitsahaliakalninaxiu2020} study local regressions and integrated coefficients. Their observation and process assumptions differ from the regional-mean experiment here. Integrated observations have their own information properties \parencite{glotergobet2008}. 

The consistency result below allows persistent covariance error, subject to an upper bound on relative variances and a separate restriction on confounding. The normal-inference results use a stronger shrinking two-sided bound directly, with the same working matrix throughout. The econometrician supplies that matrix using geography and any conditioned-on regressor information, and estimates only its variance scale from outcomes. Estimating covariance-shape parameters from the same outcome errors remains a separate problem; fixed-domain data may not identify every covariance parameter separately \parencite{zhang2004}.

\section{Regional observations and assumptions}\label{sec:model}
\subsection{Model, target, and observed averages}
Let $T\subset\R^d$ be compact and let $\mu$ be a known nonatomic Borel probability measure on $T$. The outcome, treatment, controls, and error are jointly measurable spatial random functions, defined on one probability space and unchanged as the observation resolution varies. The structural model, holding for $\mu$-almost every $s$ with probability one, is
\begin{equation}
Y(s)=\beta X(s)+W(s)'\gamma+\varepsilon(s)
\label{eq:model}
\end{equation}
The target is $\beta\in\R$. The coefficient vector $\gamma\in\R^q$ is common across locations, with a fixed finite integer $q\geq0$. An included intercept is one of the control columns. Each control is a square-integrable spatial function; it may be discontinuous or constant on subregions. When $q=0$, omit all terms involving $W$ and its coefficients.

Let $\{\mathcal P_n:n\in\mathbb N\}$ be a deterministic sequence of finite Borel partitions of $T$. Write $N_n=|\mathcal P_n|$. For each $n$, distinct regions are disjoint up to $\mu$-null sets, their union has $\mu$-measure one, and $\mu(R)>0$ for every $R\in\mathcal P_n$. For $f\in L^2(T,\mu)$, its regional mean is
\begin{equation}
f_R=\frac1{\mu(R)}\int_R f(s)\,d\mu(s)
\label{eq:averages}
\end{equation}
The econometrician observes one partition $\mathcal P=\mathcal P_n$ and its regional means $\{(Y_R,X_R,W_R):R\in\mathcal P\}$. The sequence describes the asymptotic experiment. For every observed region the averaged equation is exactly
\[
Y_R=\beta X_R+W_R'\gamma+\varepsilon_R
\]
Stack these means in $Y_n,X_n\in\R^{N_n}$ and $W_n\in\R^{N_n\times q}$. The vector $\varepsilon_n$ contains the unobserved regional errors. Regional averaging introduces no approximation to the constant-coefficient equation.

Each result is stated under the conditional law given the complete $X,W$, on any regressor realization satisfying its listed conditions. All limits below are as $n\to\infty$. Constants and fixed coefficient vectors appearing in the assumptions may depend on that regressor realization, but not on $n$ or on the remaining outcome disturbance. If the conditions hold for almost every regressor realization, so do the corresponding conditional conclusions.

\Needspace{8\baselineskip}
\begin{regularity}[Regular partitions, finite second moments, and rank condition.]\label{reg:maps}
The deterministic partitions above satisfy $N_n\to\infty$ and, for every $h>0$,
\[
\lim_{n\to\infty}
\sum_{R\in\mathcal P_n}\mu(R)
\mathbf1\{\operatorname{diam}(R)>h\}=0
\]
where $\operatorname{diam}(R)=\sup_{s,t\in R}\|s-t\|$ in the Euclidean norm. The fixed regressor realization and conditional error law satisfy
\[
\int_T\bigl\{|X(s)|^2+\|W(s)\|^2\bigr\}\,d\mu(s)<\infty
\qquad
\E\!\left[\int_T|\varepsilon(s)|^2\,d\mu(s)\,\middle|\,X,W\right]<\infty
\]
After redundant control columns are removed, there exists $n_0\in\mathbb N$ such that
\[
\operatorname{rank}[X_n\ W_n]=q+1
\qquad\text{for every }n\geq n_0
\]
The retained controls are the same $q$ spatial functions at every resolution.\statementend
\end{regularity}
The first part states that as the index of the partitions $n\to\infty$, there are increasingly many regions, and the total measure of regions whose diameter exceeds any fixed positive threshold tends to zero. The second part of the condition---full rank and square integrability---are fairly straightforward and standard in econometric practice. 

\subsection{The working kernel, true covariance, and information}
Using the conditional law specified above, define the conditional mean of the error given the complete regressor surfaces and its regional means by
\begin{equation}
m(s)=\E\{\varepsilon(s)\mid X,W\}
\qquad m_n=\E(\varepsilon_n\mid X,W)
\label{eq:conditional-mean}
\end{equation}
The entry $(m_n)_R$ is the regional average of $m(s)$ using \eqref{eq:averages}. Define the true conditional covariance separately:
\begin{equation}
\Sigma_n=\Cov(\varepsilon_n\mid X,W)
\label{eq:true-covariance}
\end{equation}
Thus $m_n$ is the conditional mean of the regional errors, and $\Sigma_n$ describes their variation around that mean. Both are unobserved model quantities. 

For every $n$, the researcher supplies/chooses an $N_n\times N_n$ matrix $K_n$ that is a measurable function of the complete $X,W$ and deterministic geography, and satisfies symmetry and positive definiteness:
\[
K_n=K_n'
\qquad
v'K_nv>0\quad\text{for every }v\in\R^{N_n}\setminus\{0\}
\]
Thus $K_n$ is fixed under the conditional outcome law. It may depend on geography and the conditioned-on regressors, but its selection cannot use the current outcome disturbances. The working specifications may change with resolution. We will discuss later whether $K_n$ must be the same as, or close to, $\Sigma_n$.

One possible construction uses a Mercer kernel $\kappa_n:T\times T\to\R$ with $\int_T\kappa_n(s,s)\,d\mu(s)<\infty$, and sets
\begin{equation}
(K_n)_{RS}=\frac1{\mu(R)\mu(S)}
\int_R\int_S \kappa_n(s,t)\,d\mu(t)\,d\mu(s)
\label{eq:kernel}
\end{equation}
This can be thought of as taking the $L^2(T,\mu)$ inner product of $\chi_R/\mu(R)$ with the Mercer kernel operator applied to $\chi_S/\mu(S)$. The chosen kernel must yield a regional matrix satisfying the strict positive-definiteness requirement above.

For a vector $f_n$ of regional values, the squared norm weighted by the inverse working covariance is
\[
f_n'K_n^{-1}f_n=\|K_n^{-1/2}f_n\|^2
\]
Here $K_n^{-1/2}$ is the symmetric inverse square root. The calculation first applies this transformation to the regional values and then sums their squares, accounting for the working variances and correlations.

Measure treatment information after controls using this quadratic form:
\begin{equation}
Q_n=\min_{\ell\in\R^q}(X_n-W_n\ell)'K_n^{-1}(X_n-W_n\ell)
\label{eq:information}
\end{equation}
$Q_n$ is the minimized sum of squared residuals from an ordinary least-squares regression of $K_n^{-1/2}X_n$ on $K_n^{-1/2}W_n$. Equivalently, it is the squared norm of the treatment residuals from GLS on $W_n$, weighted by $K_n^{-1}$.

\Needspace{6\baselineskip}
\begin{regularity}[Increasing treatment information]\label{reg:information}
For $Q_n$ defined in \eqref{eq:information}, on the fixed regressor realization,
\[
\lim_{n\to\infty}Q_n=+\infty
\]\statementend
\end{regularity}
This is the first substantive assumption: after partialling out $W$, the squared norm of the residual treatment vector, weighted by $K_n^{-1}$, must increase without bound. For fixed covariance models, this relates to the heuristic that $X(s)$ cannot be ``too smooth'' relative to the specified covariance; \textcite{dattastein2026} establish relative-smoothness conditions for consistent estimability in their Gaussian random-field models. The restrictions on the conditional error mean below separately control confounding.

While it may seem like a strong assumption, it describes an asymptotic experiment and cannot be verified directly from a single finite dataset. The practically important takeaway from \cref{reg:information} is that there must be sufficient residual treatment variation, weighted by $K_n^{-1}$, \textit{at the level at which data is observed}. Whether the limiting distribution provides a useful approximation at that resolution also depends on the covariance and conditional-mean restrictions below. The divergence of $Q_n$ is essential to the theorem, even though it cannot be established from finitely many observed partitions.

A related analogy comes from infill estimation of continuous-time processes. In a continuous semimartingale regression $Y_t=\beta X_t+\varepsilon_t$ with absolutely continuous quadratic covariations, sufficient conditions for identifying $\beta$ from fine-scale increments are
\[
    \frac{d[X,\varepsilon]_t}{dt}=0
    \qquad
    \frac{d[X,X]_t}{dt}>0
    \quad\text{for almost every }t
\]
These conditions give $[X,Y]_t=\beta[X,X]_t$, with positive treatment quadratic variation over a positive observation interval. Continuous-time models can have well-defined quadratic variation while the researcher observes only finitely many measurements, including in high-frequency financial data. These conditions describe the local variation and orthogonality on which that model's identification argument relies; their empirical usefulness depends on whether the model approximates the data well at the observed scales.

Finally, we note that while $X$ (the treatment variable on which we are conducting inference) needs sufficient residual variation, no separate information-divergence condition is imposed on the controls $W$. For example, $W$ can be $1$ on a contiguous part of the domain, and $0$ outside it, with no variation except on a boundary of measure zero. The results do not require the coefficients on $W$ to be consistently estimable, provided the rank and treatment-information conditions hold. Without further restrictions on the conditional error mean, these structural coefficients need not be separately identified from its component in the span of $W$.

\subsection{Conditions for consistency}
\begin{assumption}[True noise is bounded relative to the working covariance]\label{ass:upper}
There exist $C\in(0,\infty)$ and $n_0\in\mathbb N$ such that, for every $n\geq n_0$, the true covariance in \eqref{eq:true-covariance} and working matrix described with \eqref{eq:kernel} satisfy
\begin{equation}
0\preceq\Sigma_n\preceq C K_n
\label{eq:covariance-upper}
\end{equation}
Precisely, $0\leq v'\Sigma_n v\leq C v'K_n v$ for every $v\in\R^{N_n}$. The same finite $C$ applies at every sufficiently fine resolution.\statementend
\end{assumption}
Re-arranged, this is saying that for any vector $v$, $v'(CK_n-\Sigma_n)v\geq 0$. In other words, the working covariance $K_n$ may remain misspecified, but the true variance of every linear combination of regional errors must be at most $C$ times its working variance, with the same finite bound at all sufficiently fine resolutions. The constant $C$ is an upper bound, distinct from the scale parameter $\sigma^2$ introduced below. Compare this with weak-$*$ convergence of finite measures on a compact domain, which requires, for every continuous test function $f$,
\[
    \int f\,d(\xi_n-\xi)\longrightarrow 0
\]
The inequality in \cref{ass:upper} expresses domination at each resolution and permits persistent covariance discrepancies.

\Needspace{10\baselineskip}
\begin{assumption}[Conditional error mean after controls is small relative to treatment information]\label{ass:weak-mean}
There exists one vector $c\in\R^q$, used at every resolution, such that
\begin{equation}
\lim_{n\to\infty}
\frac{(m_n-W_nc)'K_n^{-1}(m_n-W_nc)}{Q_n}=0
\label{eq:relative-mean}
\end{equation}
Here $m_n=\E(\varepsilon_n\mid X,W)$ is defined in \eqref{eq:conditional-mean}, $K_n$ is the supplied working matrix, and $Q_n$ is defined in \eqref{eq:information}. The vector $c$ is fixed under the conditional outcome law.\statementend
\end{assumption}
For one coefficient vector $c$ used at every resolution, form $m_n-W_nc$. The squared norm of this vector after multiplication by $K_n^{-1/2}$ must grow more slowly than the treatment information $Q_n$ in \eqref{eq:information}. The condition rules out persistent confounding that reproduces a fixed nonzero multiple of the identifying treatment variation.

\subsection{Stronger assumptions for normal confidence intervals}
The preceding conditions establish consistency only. The following covariance model and stronger conditional-mean restriction justify the residual-based confidence interval in Section~\ref{sec:inference}.

\Needspace{12\baselineskip}
\begin{assumption}[Conditional Gaussian errors and asymptotically proportional covariance]\label{ass:covariance}
The conditional law of the fixed centered field $\varepsilon-m$ is Gaussian: for every finite collection $h_1,\ldots,h_J\in L^2(T,\mu)$, the vector
\[
\left(\int_T h_j(s)\{\varepsilon(s)-m(s)\}\,d\mu(s)\right)_{j=1}^{J}
\]
is jointly Gaussian conditional on $X,W$. For every $h\in L^2(T,\mu)$ with $\int_T h(s)^2\,d\mu(s)>0$,
\[
\Var\!\left(\int_T h(s)\{\varepsilon(s)-m(s)\}\,d\mu(s)
       \,\middle|\,X,W\right)>0
\]
There exist an unknown constant $\sigma^2\in(0,\infty)$, an integer $n_0$, and a sequence $(\delta_n)_{n\geq n_0}\subset[0,1)$ with $\delta_n\to0$, such that for every $n\geq n_0$,
\begin{equation}
\frac{\sigma^2}{1+\delta_n}K_n
\preceq\Sigma_n\preceq
\frac{\sigma^2}{1-\delta_n}K_n
\qquad \delta_n\longrightarrow0
\label{eq:covariance}
\end{equation}
The true covariance $\Sigma_n$ is defined in \eqref{eq:true-covariance}; $K_n$ is the actual working matrix described with \eqref{eq:kernel}. The scale and accuracy bounds are fixed under the conditional outcome law. The same error field, conditional mean, and positive scale are used across resolutions.\statementend
\end{assumption}
This assumption requires uniform relative accuracy of the scaled working covariance $\sigma^2K_n$ and eventually implies \cref{ass:upper}. For every nonzero vector $v$ of regional weights, the linear combination $v'\varepsilon_n$ has true conditional variance $v'\Sigma_nv$ and scaled working variance $\sigma^2v'K_nv$. These two variances must differ by at most a proportion $\delta_n$ of the true variance, uniformly over $v$. Equivalently,
\[
\sup_{v\ne0}
\left|\frac{\sigma^2v'K_nv}{v'\Sigma_nv}-1\right|\leq\delta_n
\]
The requirement concerns all linear combinations of regional errors, rather than matrix entries individually. The case $\delta_n=0$ is exact proportionality. Otherwise the working matrix may be wrong at every finite resolution. The scale $\sigma^2$ is estimated from residuals.

\Needspace{9\baselineskip}
\begin{assumption}[A fixed conditional error mean with a bounded quadratic form after controls]\label{ass:mean}
Let $m(s)=\E\{\varepsilon(s)\mid X,W\}$ be the fixed conditional-mean function in \eqref{eq:conditional-mean}. There exists one vector $c\in\R^q$, used at every resolution, such that
\begin{equation}
\sup_{n\geq1}(m_n-W_nc)'K_n^{-1}(m_n-W_nc)<\infty
\label{eq:mean-energy}
\end{equation}
In this expression $m_n-W_nc$ consists of regional averages of the single function $m-W'c\in L^2(T,\mu)$. Neither that function nor $c$ may be reselected at each resolution. The supremum is a finite number under the conditional regressor realization.\statementend
\end{assumption}
The vector $m_n-W_nc$ consists of regional averages of the same spatial function $m-W'c$ at every resolution. The assumption bounds the squared norm $\|K_n^{-1/2}(m_n-W_nc)\|^2$ uniformly over $n$. Together with $Q_n\to\infty$, this implies the consistency condition \eqref{eq:relative-mean}. Under the preceding refinement and covariance assumptions, Appendix~\ref{app:mean} proves the stronger conclusion that the conditional slope bias is negligible relative to $Q_n^{-1/2}$. The function $m-W'c$ itself may remain nonzero. 

This is a sufficient restriction on the conditional error mean for local exogeneity. It concerns only the one fixed mean offset $m-W'c$ stipulated above; neither the function nor $c$ may vary with resolution. The restriction concerns unobserved omitted determinants and requires substantive justification.

\section{GLS estimation, results, and discussion}\label{sec:inference}
\subsection{One coefficient regression and one residual-scale calculation}
Fit the complete regression using the working matrix $K_n$ from \eqref{eq:kernel}:
\begin{equation}
(\widehat\beta_n,\widehat\gamma_n)
=\argmin_{b,g}(Y_n-bX_n-W_ng)'K_n^{-1}(Y_n-bX_n-W_ng)
\label{eq:estimator}
\end{equation}
All included coefficients are estimated. Its equivalent residualization is
\begin{equation}
x_n=K_n^{-1/2}
\left[X_n-W_n(W_n'K_n^{-1}W_n)^{-1}W_n'K_n^{-1}X_n\right]
\label{eq:x}
\end{equation}
\begin{equation}
y_n=K_n^{-1/2}
\left[Y_n-W_n(W_n'K_n^{-1}W_n)^{-1}W_n'K_n^{-1}Y_n\right]
\label{eq:y}
\end{equation}
\begin{equation}
\widehat\beta_n=\frac{x_n'y_n}{x_n'x_n}
\qquad Q_n=x_n'x_n
\label{eq:FWL}
\end{equation}
Use the symmetric inverse square root and omit the control terms when there are no controls. The bracketed terms subtract fitted control contributions; multiplication by $K_n^{-1/2}$ transforms them so that the GLS criterion is an ordinary sum of squares. These are full-regression Frisch--Waugh--Lovell identities \parencite{ding2021}. The information $Q_n$ is defined in \eqref{eq:information}.

Using the same fitted regression, calculate
\begin{equation}
\widehat\varepsilon_n=Y_n-X_n\widehat\beta_n-W_n\widehat\gamma_n
\label{eq:residuals}
\end{equation}
\begin{equation}
\widehat\sigma_n^2
=\frac{\widehat\varepsilon_n'K_n^{-1}\widehat\varepsilon_n}{N_n-q-1}
\qquad
\widehat\Sigma_n=\widehat\sigma_n^2K_n
\label{eq:scale}
\end{equation}
The divisor removes the treatment and all control degrees of freedom, including an intercept when present. Under \Cref{ass:covariance}, the residual quadratic form is positive almost surely once $N_n>q+1$. Use any fixed positive fallback at earlier undefined resolutions. Multiplying a covariance matrix by a positive scalar leaves the GLS coefficients unchanged, so the computation needs no second coefficient regression.

The reported variance is
\begin{equation}
\boxed{\widehat V_{\beta,n}=\frac{\widehat\sigma_n^2}{Q_n}
=\frac{\widehat\sigma_n^2}{x_n'x_n}}
\label{eq:variance}
\end{equation}
It equals the treatment entry of $\{[X_n\ W_n]'\widehat\Sigma_n^{-1}[X_n\ W_n]\}^{-1}$ and retains estimated controls. Under only the consistency assumptions, \eqref{eq:scale} is a computable residual summary with no necessary common population scale. Under the shrinking bounds in \Cref{ass:covariance} and the fixed-mean condition in \Cref{ass:mean}, Proposition~\ref{prop:variance} proves that it estimates $\sigma^2$ and that \eqref{eq:variance} estimates the actual slope variance. The same $K_n$ is used in every calculation.

\subsection{Three propositions}
All limits condition on the complete $X,W$. The true covariance $\Sigma_n$, working matrix $K_n$, and information $Q_n$ are defined in \eqref{eq:true-covariance}, \eqref{eq:kernel}, and \eqref{eq:information}, respectively.

\Needspace{10\baselineskip}
\begin{proposition}[Consistency of GLS under covariance misspecification]\label{prop:consistency}
Under \Cref{reg:maps,reg:information,ass:upper,ass:weak-mean}, the estimator in \eqref{eq:estimator} satisfies
\[
\widehat\beta_n\pto\beta
\]
At every sufficiently fine resolution, for the constant $C$ and the fixed control coefficient vector $c$ in \Cref{ass:upper,ass:weak-mean},
\begin{equation}
\E\{(\widehat\beta_n-\beta)^2\mid X,W\}
\leq\frac{C+(m_n-W_nc)'K_n^{-1}(m_n-W_nc)}{Q_n}
\longrightarrow0
\label{eq:mse-bound}
\end{equation}
\end{proposition}
\proofref{app:coefficient-proofs}

\Needspace{8\baselineskip}
\begin{proposition}[Asymptotic normality of GLS]\label{prop:normality}
Under \Cref{reg:maps,reg:information,ass:covariance,ass:mean}, with the scale $\sigma^2$ in \eqref{eq:covariance},
\begin{equation}
\frac{\sqrt{Q_n}}{\sigma}(\widehat\beta_n-\beta)
\dto\mathcal N(0,1)
\label{eq:normality}
\end{equation}
\end{proposition}
\proofref{app:coefficient-proofs}
This proposition uses \Cref{ass:covariance,ass:mean} in addition to the observation and information conditions. The true conditional slope variance is asymptotic to $\sigma^2/Q_n$, although it need not equal that expression at a finite resolution.

\Needspace{12\baselineskip}
\begin{proposition}[Variance estimation and confidence intervals]\label{prop:variance}
Under \Cref{reg:maps,reg:information,ass:covariance,ass:mean}, the residual scale in \eqref{eq:scale} and reported variance in \eqref{eq:variance} satisfy
\begin{equation}
\widehat\sigma_n^2\pto\sigma^2
\qquad
\frac{\widehat V_{\beta,n}}{\Var(\widehat\beta_n\mid X,W)}\pto1
\label{eq:variance-consistency}
\end{equation}
Also $\widehat V_{\beta,n}/(\sigma^2/Q_n)\pto1$. Consequently,
\[
\frac{\widehat\beta_n-\beta}{\sqrt{\widehat V_{\beta,n}}}
\dto\mathcal N(0,1)
\]
and $\widehat\beta_n\pm z_{1-\alpha/2}\sqrt{\widehat V_{\beta,n}}$ has limiting conditional coverage $1-\alpha$, where $z_{1-\alpha/2}$ is the standard-normal quantile.
\end{proposition}
\proofref{app:variance-proof}

\section{U.S. geography simulation study}\label{sec:simulation}

We implement the estimator with the accompanying \texttt{infillgls} R
package on the 2024 U.S.~Census Cartographic Boundary Files.  The corrected design observes every
retained region: all 50 states and the District of Columbia ($51$ regions) and
all $3{,}144$ county-equivalents.  There is no region sampling,
stratification, or within-region location sampling.  U.S.~territories are
outside the retained study domain.  A dense covariance calculation over all
census tracts is outside the computational scope of this implementation, so
tracts are explicitly excluded; this study makes no tract simulation, map, or
coverage claim.

For every retained polygon, we calculate one deterministic point on its
surface after projection to EPSG:6933 and transform it back to geographic
coordinates.  The covariance matrices are exact dense point-support
covariances for those deterministic representative points.  Thus these are
not sampled locations and are not exact polygon-integral covariances.  For visualization, we draw one shared,
zero-nugget Mat\'ern-$1$ latent surface at all $3{,}195$ regional supports
($3{,}193$ unique locations, since two state/county supports coincide);
\Cref{fig:us-geometry-surface} displays the value at each representative
point.

The data-generating point covariance is Mat\'ern-$1$ using Haversine distance,
range $700$ km, marginal variance $1$, and point-level nugget $0.05$.  At
each level, a fixed, seeded geographically varying treatment and standardized
longitude and latitude controls are held constant across replications; the
true treatment coefficient is one.  We generate $500$ outcome-surface draws
per scenario and level and fit \texttt{spatial\_gls}.  Besides the matched Mat\'ern-$1$ matrix, we
use a slight Mat\'ern change ($\nu=0.85$, range $630$ km), precisely the same
slight matrix multiplied by $C=1.75$, and a deliberately severe compound
covariance-specification mismatch: a squared-exponential working covariance
based on three-dimensional Earth-chord distance with range $350$ km.  

\begin{table}[H]
\centering
\small
\caption{Empirical nominal 95\% interval coverage over 500 simulated outcome
surfaces per scenario and geographic level, using every retained state/DC and
county-equivalent region.  The final column is a binomial 95\% Wilson Monte
Carlo interval for the reported coverage proportion.}
\label{tab:us-geometry-coverage}
\input{us_geography_all_regions_simulation_summary.tex}

\end{table}

At the state/DC level, matched working covariance coverage is $0.942$
(Wilson interval $[0.918,0.959]$), and the slight shape error gives $0.944$
($[0.920,0.961]$); both Monte Carlo intervals contain the nominal value.
The compound severe mismatch undercovers at this level ($0.850$;
$[0.816,0.879]$).  At the county level, coverage is $0.962$ for the matched
covariance and $0.958$ for each of the slight-shape and compound-severe
covariances; all four county Wilson intervals contain $0.95$.  These scenarios
do not isolate the effect of covariance functional form: the severe case
jointly changes the family, distance metric, and range.  Its state/DC result
shows undercoverage for that compound mismatch, whereas its county result
($0.958$; $[0.937,0.972]$) is compatible with nominal coverage in this
particular all-county finite design.  Scaling the slight working covariance by
$C=1.75$ leaves the paired point estimates, standard errors, p-values, and
confidence intervals unchanged to a tolerance of $10^{-10}$; $Q_n$ and the
residual-scale estimate change by compensating factors.  

\section{Conclusion}
This article covers consistency, asymptotic normality, and valid inference for spatial generalized least squares under infill asymptotics. The conditions for locally exogenous variation combine increasing treatment information after controls with restrictions that make the conditional bias of the GLS slope vanish as resolution increases. They allow treatment to be endogenous in levels and errors to remain spatially correlated. The stronger inference conditions make the bias negligible relative to the slope's standard deviation.
\printbibliography[heading=bibintoc,title={References}]

\appendix
\section{Notation}\label{app:notation}
Definitions and their locations are listed below. The GLS outcome-weight vector $\omega_n$ is introduced only in the proofs; it is the vector multiplying outcomes to produce the estimated slope.
\begingroup\small
\renewcommand{\arraystretch}{1.12}
\begin{longtable}{@{}>{\raggedright\arraybackslash}p{.19\textwidth}>{\raggedright\arraybackslash}p{.55\textwidth}>{\raggedright\arraybackslash}p{.20\textwidth}@{}}
\toprule Symbol & Meaning and availability & Defined in \\\midrule\endfirsthead
\toprule Symbol & Meaning and availability & Defined in \\\midrule\endhead
\bottomrule\endfoot
$T,\mu,\mathcal P_n$ & Fixed domain, observation measure, and deterministic partition & Section~\ref{sec:model}\\
$N_n,q,f_R$ & Number of observed regions, control columns, and regional average & \eqref{eq:averages}\\
$X_n,Y_n,W_n$ & Observed regional treatment, outcome, and controls & Section~\ref{sec:model}\\
$K_n,\kappa_n,\kappa$ & GLS working matrix, its optional resolution-dependent kernel function, and the fixed-kernel special case & \eqref{eq:kernel}\\
$m,m_n$ & Conditional error mean function and its regional means; unobserved & \eqref{eq:conditional-mean}\\
$\Sigma_n$ & True conditional regional-error covariance; unobserved & \eqref{eq:true-covariance}\\
$Q_n$ & Treatment information after controls, weighted by $K_n^{-1}$; computed & \eqref{eq:information}\\
$\ell,c$ & Dummy treatment-on-controls projection coefficient; fixed coefficients used to subtract $W_nc$ from $m_n$ in the mean restrictions & \eqref{eq:information}; \eqref{eq:relative-mean}, \eqref{eq:mean-energy}\\
$C,C_0$ & Finite covariance upper bound and optional positive lower bound & \eqref{eq:covariance-upper}, \eqref{eq:two-sided}\\
$\delta_n$ & Vanishing relative covariance-error bound\\
$\sigma^2$ & Unknown limiting covariance scale in the shrinking two-sided bound & \eqref{eq:covariance}\\
$x_n,y_n$ & Fitted control residuals after multiplication by $K_n^{-1/2}$ & \eqref{eq:x}--\eqref{eq:y}\\
$\widehat\varepsilon_n$ & Complete regression residuals & \eqref{eq:residuals}\\
$\widehat\sigma_n^2,\widehat\Sigma_n$ & Residual-scale estimate and fitted scale times the working matrix & \eqref{eq:scale}\\
$\widehat V_{\beta,n}$ & Reported coefficient variance & \eqref{eq:variance}\\
$\omega_n$ & GLS outcome-weight vector at $K_n$; proof notation & \eqref{eq:weights}\\
$\Pi_n$ & Regional-averaging projection; proof notation & Lemma~\ref{lem:averaging}\\
\end{longtable}
\endgroup

\section{Proofs of the three propositions}\label{app:proofs}
Throughout, condition on a fixed complete regressor realization satisfying Section~\ref{sec:model}. All matrices act on the full set of regional observations. The true covariance $\Sigma_n$ is defined in \eqref{eq:true-covariance}; the working matrix $K_n$ and information $Q_n$ are in \eqref{eq:kernel}--\eqref{eq:information}. Constants may depend on the fixed regressor realization and working kernel. Each proof states which covariance assumptions it uses.

\subsection{Regional means and the GLS outcome weights}\label{app:averaging}
\begin{lemma}[Refinement in measure recovers square-integrable functions]\label{lem:averaging}
Let $\Pi_n f$ equal $f_R$ on each $R\in\mathcal P_n$, where $f_R$ is defined in \eqref{eq:averages}. Under the refinement condition in \Cref{reg:maps},
\[
\|\Pi_n f-f\|_{L^2(T,\mu)}\longrightarrow0
\qquad\text{for every }f\in L^2(T,\mu)
\]
\end{lemma}
\begin{proof}
Regional averaging is an orthogonal projection, since its residual integrates to zero on every region. Suppose first that $f$ is continuous on compact $T$. For every $h>0$,
\[
\begin{split}
\|\Pi_n f-f\|_{L^2(T,\mu)}^2
\leq{}&\sup_{\substack{s,t\in T\\\|s-t\|\leq h}}|f(s)-f(t)|^2\\
&+4\|f\|_\infty^2\sum_{R\in\mathcal P_n}
          \mu(R)\mathbf1\{\operatorname{diam}(R)>h\}
\end{split}
\]
The second term tends to zero for fixed $h$ by \Cref{reg:maps}. Uniform continuity makes the first term tend to zero as $h\downarrow0$. Taking these limits in that order gives convergence in spatial $L^2$. Continuous functions are dense in $L^2$ for a finite Borel measure on compact $T$, and each $\Pi_n$ is an $L^2$ contraction. Approximation therefore extends the conclusion to every square-integrable $f$. The argument permits nonnested partitions and does not require the largest region diameter to vanish.
\end{proof}

For the proofs, the vector of GLS outcome weights at the specified $K_n$ is
\begin{equation}
\omega_n=
\frac{K_n^{-1}\{X_n-W_n(W_n'K_n^{-1}W_n)^{-1}W_n'K_n^{-1}X_n\}}{Q_n}
\label{eq:weights}
\end{equation}
The denominator $Q_n$ is defined in \eqref{eq:information}. Direct regression algebra gives
\begin{equation}
\widehat\beta_n=\omega_n'Y_n
\qquad \omega_n'X_n=1
\qquad \omega_n'W_n=0
\qquad \omega_n'K_n\omega_n=Q_n^{-1}
\label{eq:weight-identities}
\end{equation}
The weights multiply the outcomes to obtain the slope; they are not fitted treatment values. Their definition contains only the regional regressors and the working matrix specified with \eqref{eq:kernel}.

\subsection{Why the conditional error mean has a negligible effect on the slope}\label{app:mean}
The true covariance matrices $\Sigma_n$ come from regional averages of the same centered error field at every resolution, while the working matrices $K_n$ may change with resolution. The following argument first studies GLS using these true covariance matrices, then bounds the change caused by using the working weights. All additional notation in this proof is confined to the appendix.

\begin{lemma}[The stipulated conditional mean has a negligible slope contribution]\label{lem:mean}
Maintain \Cref{reg:maps,reg:information,ass:covariance,ass:mean}. Then the working outcome weights in \eqref{eq:weights} satisfy
\begin{equation}
\sqrt{Q_n}\,\omega_n'm_n\longrightarrow0.
\label{eq:bias}
\end{equation}
\end{lemma}
\begin{proof}
Let $c$ be the single vector from \Cref{ass:mean}, and write $f=m-W'c$, so that $f_n=m_n-W_nc$. Write $u=\varepsilon-m$ and $D_n=[X_n\ W_n]$ in this proof. For the regression using the true covariance, define the coefficient-covariance matrix, its treatment variance, and its outcome-weight vector by
\[
B_n=(D_n'\Sigma_n^{-1}D_n)^{-1}
\qquad v_n=(B_n)_{11}
\qquad \omega_n^*=\Sigma_n^{-1}D_nB_ne_1
\]
where $e_1$ selects the treatment coefficient. These quantities describe hypothetical GLS using the true covariance and are introduced only for the proof. The bounds in \eqref{eq:covariance} imply
\begin{equation}
(1-\delta_n)\Sigma_n\preceq\sigma^2K_n\preceq(1+\delta_n)\Sigma_n
\label{eq:proof-relative-order}
\end{equation}
They also imply
\[
f_n'\Sigma_n^{-1}f_n
\leq\frac{1+\delta_n}{\sigma^2}f_n'K_n^{-1}f_n
\qquad
\frac{\sigma^2}{(1+\delta_n)Q_n}
\leq v_n\leq\frac{\sigma^2}{(1-\delta_n)Q_n}
\]
The variance bounds follow by minimizing the respective covariance quadratic forms over $D_n'w=e_1$. Hence $v_n\to0$, $Q_nv_n\to\sigma^2$, and $f_n'\Sigma_n^{-1}f_n$ is uniformly bounded. The reverse quadratic-form inequality follows from the other side of \eqref{eq:proof-relative-order}.

\emph{GLS using the true covariance of the fixed field.}
Let $\mathcal H$ be the closed linear span, in the Hilbert space of square-integrable random variables with conditional inner product $\E(UV\mid X,W)$, of spatial averages $\int_Tu(s)h(s)\,d\mu(s)$, $h\in L^2(T,\mu)$. The finite conditional second moment $\E[\int_Tu(s)^2\,d\mu(s)\mid X,W]<\infty$ makes $V\mapsto\E\{u(\cdot)V\mid X,W\}$ a map into spatial $L^2$. Its image, with the inherited norm, is the covariance function space.

The bound on this fixed $f_n'\Sigma_n^{-1}f_n$ gives $|t'f_n|\leq C(t'\Sigma_nt)^{1/2}$ for every regional coefficient vector $t$. For fixed $h$, choose $t_R=\int_Rh\,d\mu$. The corresponding spatial averaging function is $\Pi_nh$, defined in \Cref{lem:averaging}. By \Cref{lem:averaging}, $\Pi_nh\to h$ in spatial $L^2$; both the mean integral and its error variance converge. Riesz representation therefore gives one fixed $V\in\mathcal H$ such that $f(s)=\E\{u(s)V\mid X,W\}$.

Form the nested common refinements of the first $k$ observed partitions, omitting zero-measure cells. Their true covariance matrices are positive definite by the spatial-average positivity in \Cref{ass:covariance}. Once their designs have full rank, their inverse information matrices decrease in positive-semidefinite order to a matrix $B$. Each observed experiment is a linear image of a sufficiently fine common refinement, so $B_n\succeq B$.

Fix one such common-refinement experiment. Project its finitely many averaging functions onto the observed partition at level $n$. The resulting linear combinations of regional observations have design and covariance matrices converging to those of the fixed experiment, by \Cref{lem:averaging} and the finite conditional second moment in spatial $L^2$. Their covariance and information matrices are nonsingular eventually. Full GLS has no larger coefficient covariance than GLS on these combinations. Take the limit superior in every quadratic form as $n\to\infty$, then refine the fixed experiment. Together with $B_n\succeq B$, this proves $B_n\to B$.

Since $v_n\to0$, the first row and column of $B$ vanish. Positivity of $B_n-B$ gives
\[
\frac{|(B_n)_{1j}|}{\sqrt{v_n}}
\leq\sqrt{(B_n-B)_{jj}}\longrightarrow0
\]
for every fixed coefficient coordinate $j$. The random variable $Z_n=(\omega_n^*)'u_n/\sqrt{v_n}$ has conditional $L^2$ norm one. For $t_R=\int_Rh\,d\mu$,
\[
\Cov\left\{Z_n,\int_Tu(s)(\Pi_nh)(s)\,d\mu(s)\mid X,W\right\}
=\frac{e_1'B_nD_n't}{\sqrt{v_n}}\longrightarrow0
\]
Here $D_n't$ converges to the spatial inner products of $h$ with the fixed regressors. Replacing $\Pi_nh$ by $h$ changes the covariance by a vanishing amount. These averages are dense in $\mathcal H$, so $Z_n$ converges weakly to zero in that Hilbert space:
\[
\E(Z_nV\mid X,W)\longrightarrow0
\qquad\text{for every fixed }V\in\mathcal H
\]
Each $Z_n$ remains conditionally standard normal; the limit above describes its inner products with fixed elements of $\mathcal H$. Taking $V$ to be the fixed representer of $f$ yields
\[
\frac{(\omega_n^*)'f_n}{\sqrt{v_n}}\longrightarrow0
\]
This part of the argument uses that the regional errors at every resolution are averages of the same fixed field, and does not require the observed partitions themselves to be nested.

\emph{Transfer to the working weights.}
Suppress subscripts only in this paragraph, so $K$ denotes the working matrix $K_n$. Both $\omega^*$ and $\omega$ minimize a positive-definite quadratic form under the constraint $D'w=e_1$: their matrices are $\Sigma$ and $\sigma^2 K$, respectively. Define the proof-only vectors and relative-error matrix
\[
a=\Sigma^{1/2}\omega^*
\qquad d=\Sigma^{1/2}(\omega-\omega^*)
\qquad E=\Sigma^{-1/2}(\sigma^2K-\Sigma)\Sigma^{-1/2}
\]
The covariance bounds give $\|E\|_{\mathrm{op}}\leq\delta$. The first-order conditions give $d'a=0$ and $d'(I+E)(a+d)=0$. Therefore
\[
(1-\delta)\|d\|^2\leq-d'Ea
\leq\delta\|d\|\|a\|
\]
Since $\|a\|^2=v$, division when $d\ne0$ gives, restoring subscripts,
\begin{equation}
(\omega_n-\omega_n^*)'\Sigma_n(\omega_n-\omega_n^*)
\leq\frac{\delta_n^2}{(1-\delta_n)^2}v_n
\label{eq:working-weight-perturbation}
\end{equation}
The same inequality is immediate if $d=0$. Covariance Cauchy--Schwarz now gives
\[
\sqrt{Q_n}\,|(\omega_n-\omega_n^*)'f_n|
\leq
\frac{\delta_n\sqrt{Q_nv_n}}{1-\delta_n}
\{f_n'\Sigma_n^{-1}f_n\}^{1/2}\longrightarrow0
\]
Together with the result for GLS using the true covariance and $Q_nv_n\to\sigma^2$, this proves that $\sqrt{Q_n}\,\omega_n'f_n\to0$. Since $\omega_n'W_n=0$ and $f_n=m_n-W_nc$, it proves \eqref{eq:bias}. No rate linking $\delta_n$ to $N_n$ or $Q_n$ is used.
\end{proof}

\subsection{Residual-scale consistency}\label{app:scale}
\begin{lemma}[Residuals consistently estimate the limiting variance scale]\label{lem:scale}
Under \Cref{reg:maps,ass:covariance,ass:mean}, the residual-scale estimate in \eqref{eq:scale} satisfies
\[
\widehat\sigma_n^2-\sigma^2=O(\delta_n)+O_p(N_n^{-1/2})
\]
and hence converges in probability to $\sigma^2$.
\end{lemma}
\begin{proof}
Let $P_n$ be the orthogonal residual projection after regressing in $K_n^{-1/2}$ coordinates on the complete design $K_n^{-1/2}[X_n\ W_n]$. Its rank is $N_n-q-1$. Define, only in this proof, the transformed centered covariance and the projected conditional mean by
\[
\Gamma_n=K_n^{-1/2}\Sigma_nK_n^{-1/2}
\qquad b_n=P_nK_n^{-1/2}m_n
\]
The covariance assumption \eqref{eq:covariance} gives
\[
\frac{\sigma^2}{1+\delta_n}I\preceq\Gamma_n
\preceq\frac{\sigma^2}{1-\delta_n}I
\]
Because $P_nK_n^{-1/2}W_n=0$, \eqref{eq:mean-energy} gives $\|b_n\|^2\leq C$ for some fixed finite constant. The residual numerator in \eqref{eq:scale} is the squared norm of a projected Gaussian vector with mean $b_n$. Its exact conditional expectation and variance are
\[
\tr(P_n\Gamma_n)+\|b_n\|^2
\]
and
\[
2\tr(P_n\Gamma_nP_n\Gamma_n)+4b_n'\Gamma_nb_n
\]
respectively. The trace of $P_n\Gamma_n$ lies between $\sigma^2(N_n-q-1)/(1+\delta_n)$ and $\sigma^2(N_n-q-1)/(1-\delta_n)$. Consequently
\[
\left|\E(\widehat\sigma_n^2\mid X,W)-\sigma^2\right|
\leq\frac{\sigma^2\delta_n}{1-\delta_n}
+\frac{C}{N_n-q-1}
\]
The quadratic-form variance is bounded by
\[
\frac{2\sigma^4}{(1-\delta_n)^2}(N_n-q-1)
+\frac{4\sigma^2C}{1-\delta_n}
\]
Dividing by $(N_n-q-1)^2$ gives a conditional variance of order $N_n^{-1}$. Chebyshev proves the displayed rate and consistency. The coefficient and scale may use the same outcomes; their independence is unnecessary.
\end{proof}

\subsection{Consistency and asymptotic normality}\label{app:coefficient-proofs}
\begin{proof}[Proof of \Cref{prop:consistency}]\label{proof:consistency}
Use the GLS outcome weights $\omega_n$ in \eqref{eq:weights} and their identities \eqref{eq:weight-identities}. The exact coefficient error is
\begin{equation}
\widehat\beta_n-\beta
=\omega_n'm_n+\omega_n'(\varepsilon_n-m_n)
\label{eq:decomposition}
\end{equation}
where $m_n$ is defined in \eqref{eq:conditional-mean}. Its conditional mean-squared error is exactly
\begin{equation}
\E\{(\widehat\beta_n-\beta)^2\mid X,W\}
=(\omega_n'm_n)^2+\omega_n'\Sigma_n\omega_n
\label{eq:exact-mse}
\end{equation}
The true covariance $\Sigma_n$ is defined in \eqref{eq:true-covariance}. For the fixed control vector $c$ in \Cref{ass:weak-mean}, $\omega_n'W_n=0$ and Cauchy--Schwarz give
\begin{equation}
\begin{split}
|\omega_n'm_n|^2
&=|\omega_n'(m_n-W_nc)|^2\\
&\leq (\omega_n'K_n\omega_n)
          (m_n-W_nc)'K_n^{-1}(m_n-W_nc)\\
&=\frac{(m_n-W_nc)'K_n^{-1}(m_n-W_nc)}{Q_n}
\end{split}
\label{eq:bias-bound}
\end{equation}
The upper covariance bound \eqref{eq:covariance-upper} gives
\[
\omega_n'\Sigma_n\omega_n\leq C\omega_n'K_n\omega_n=\frac{C}{Q_n}
\]
Combine these inequalities to obtain \eqref{eq:mse-bound}. The information and relative-mean conditions make its right side vanish. Markov's inequality applied to the squared coefficient error proves convergence in probability. This argument uses conditional second moments only. It does not require a lower covariance bound, Gaussianity, or the sharper bias conclusion \eqref{eq:bias}. Under \Cref{ass:covariance,ass:mean}, the upper bound holds eventually with $C=2\sigma^2$, and the bound in \eqref{eq:mean-energy} implies \eqref{eq:relative-mean}. Consistency is therefore also available under the stronger inference assumptions.
\end{proof}

\begin{proof}[Proof of \Cref{prop:normality}]\label{proof:normality}
The centered term in \eqref{eq:decomposition} is conditionally Gaussian, because the working outcome weights are fixed conditional on $X,W$. Its variance after multiplication by $\sqrt{Q_n}/\sigma$ lies between $1/(1+\delta_n)$ and $1/(1-\delta_n)$, by \eqref{eq:covariance} and \eqref{eq:weight-identities}. That variance tends to one. The scaled conditional mean tends to zero by \eqref{eq:bias}. This proves \eqref{eq:normality}. In particular the actual conditional slope variance is $\sigma^2Q_n^{-1}\{1+o(1)\}$, rather than being assumed equal to this quantity at every resolution.
\end{proof}

\subsection{Variance estimation and confidence intervals}\label{app:variance-proof}
\begin{proof}[Proof of \Cref{prop:variance}]\label{proof:variance}
\Cref{lem:scale} proves $\widehat\sigma_n^2\pto\sigma^2$. The Schur-complement identity for the complete regression gives the treatment entry of $([X_n\ W_n]'K_n^{-1}[X_n\ W_n])^{-1}$ as $1/Q_n$. The actual conditional variance is $\omega_n'\Sigma_n\omega_n$. Applying \eqref{eq:covariance} and \eqref{eq:weight-identities} yields
\[
\frac{\sigma^2}{1+\delta_n}
\leq Q_n\omega_n'\Sigma_n\omega_n
\leq\frac{\sigma^2}{1-\delta_n}
\]
Hence the ratio of the reported variance $\widehat\sigma_n^2/Q_n$ to this actual variance converges to one, as does its ratio to $\sigma^2/Q_n$. Combine with Proposition~\ref{prop:normality} and Slutsky's theorem. Continuity of the normal distribution at the interval endpoints proves limiting coverage. Exact finite-sample Student-$t$ coverage under confounding is not asserted.
\end{proof}

\subsection{Persistent covariance bounds and fixed working kernels}\label{app:misspecification}
For constants $0<C_0\leq C<\infty$, independent of resolution, a two-sided covariance bound gives
\begin{equation}
C_0K_n\preceq\Sigma_n\preceq CK_n
\quad\Longrightarrow\quad
\frac{C_0}{Q_n}\leq\Var(\widehat\beta_n\mid X,W)\leq\frac{C}{Q_n}
\label{eq:two-sided}
\end{equation}
For arbitrary fixed conditional working matrices, the identity $\omega_n'K_n\omega_n=1/Q_n$ proves \eqref{eq:two-sided}'s variance bounds. They do not equate the actual treatment variance and the residual variance estimate.

Indeed, with the residual projection $P_n$ defined in Appendix~\ref{app:scale}, the exact expectation under a general true covariance is
\begin{equation}
\E(\widehat\sigma_n^2\mid X,W)
=\frac{\tr\{P_nK_n^{-1/2}\Sigma_nK_n^{-1/2}\}
+\|P_nK_n^{-1/2}m_n\|^2}{N_n-q-1}
\label{eq:misspec-scale-mean}
\end{equation}
The target variance depends on $\omega_n'\Sigma_n\omega_n$, rather than this average of residual-direction variances. Fixed eigenvalue bounds do not equate their limits. The shrinking bound in \Cref{ass:covariance} makes all the transformed variances approach $\sigma^2$, and Lemma~\ref{lem:scale} proves the required concentration.

\paragraph{Why a fixed wrong kernel cannot become relatively correct in all directions.}
Suppose $K_n$ exactly integrates a single fixed kernel $\kappa$, and the true field is fixed. For $h\in L^2(T,\mu)$, choose $v_R=\int_Rh\,d\mu$. Then $v'\Sigma_nv$ is the variance of $\int_Tu(s)(\Pi_nh)(s)\,d\mu(s)$, while $v'K_nv$ is the corresponding variance under the working kernel. Lemma~\ref{lem:averaging} gives $\Pi_nh\to h$ in spatial $L^2$. Finite integrated variance makes both covariance forms continuous for this convergence. Taking limits in \eqref{eq:covariance} therefore equates the true variance of every such spatial average to $\sigma^2$ times its working variance. Polarization equates the two covariance operators. Thus a permanent discrepancy in the variance of any spatial average is excluded. Genuine finite-resolution approximation error is possible because the main model permits the working matrices to change beyond merely integrating one fixed kernel.

\end{document}

%% file: us_geography_all_regions_simulation_summary.tex
\begin{tabular}{llrrr}
\toprule
Working covariance & Geography & $n$ & 95\% coverage & 95\% Monte Carlo interval \\
\midrule
Matched Matern-1 & state & 51 & 0.942 & [0.918, 0.959] \\
Slight Matern shape & state & 51 & 0.944 & [0.920, 0.961] \\
Slight Matern shape, $C=1.75$ & state & 51 & 0.944 & [0.920, 0.961] \\
Severe squared-exponential & state & 51 & 0.850 & [0.816, 0.879] \\
Matched Matern-1 & county & 3144 & 0.962 & [0.941, 0.976] \\
Slight Matern shape & county & 3144 & 0.958 & [0.937, 0.972] \\
Slight Matern shape, $C=1.75$ & county & 3144 & 0.958 & [0.937, 0.972] \\
Severe squared-exponential & county & 3144 & 0.958 & [0.937, 0.972] \\
\bottomrule
\end{tabular}